\documentclass[conference]{IEEEtran}
\IEEEoverridecommandlockouts
\usepackage[noadjust]{cite}
\usepackage{amsmath,amssymb,amsfonts}
\usepackage{algorithmic}
\usepackage{graphicx}
\usepackage{textcomp}
\usepackage{listings} 
\usepackage{tabularx} 
\usepackage{booktabs} 
\usepackage{multirow} 
\usepackage{makecell} 
\usepackage{xcolor}
\def\BibTeX{{\rm B\kern-.05em{\sc i\kern-.025em b}\kern-.08em
    T\kern-.1667em\lower.7ex\hbox{E}\kern-.125emX}}
\begin{document}


\title{Can LLMs Help Predict Elections? (Counter) Evidence from the World’s Largest Democracy}

\author{
\IEEEauthorblockN{Pratik Gujral\IEEEauthorrefmark{1}, 
Kshitij Awaldhi\IEEEauthorrefmark{1}, Navya Jain\IEEEauthorrefmark{1}, Bhavuk Bhandula\IEEEauthorrefmark{1} and
Abhijnan Chakraborty\IEEEauthorrefmark{2}}
\IEEEauthorblockA{\IEEEauthorrefmark{1}Indian Institute of Technology Delhi, India}
\IEEEauthorblockA{\IEEEauthorrefmark{2}Indian Institute of Technology Kharagpur, India}
}

\maketitle

\begin{abstract}
The study of how social media affects the formation of public opinion and its influence on political results has been a 
popular field of inquiry. However, current approaches frequently offer a limited comprehension of the complex political phenomena,  
yielding inconsistent outcomes. In this work, we introduce a new method: harnessing the capabilities of Large Language Models (LLMs) to examine social media data and forecast election outcomes.
Our research diverges from traditional methodologies in two crucial respects. First, we utilize the sophisticated capabilities of foundational LLMs, which can comprehend the complex linguistic subtleties and contextual details present in social media data. 
Second, 
we focus on data from X (Twitter) in India to predict state assembly election outcomes. 
Our method entails 
sentiment analysis of election-related tweets 
through LLMs to forecast the actual election results, and we demonstrate the superiority of our LLM-based method against more traditional exit and opinion polls. 
Overall, our research offers valuable insights into the unique dynamics of Indian politics and the remarkable impact of social media in molding public attitudes within this context.
\end{abstract}

\begin{IEEEkeywords}
Large Language Models, Social Media Analysis, Sentiment Analysis, Election Prediction, Politics, X (Twitter).
\end{IEEEkeywords}

\section{Introduction}

Elections play a pivotal role in democracies, serving as the expression of collective public will, and the means for peaceful transition of power. 
Multiple research works have linked political stability with economic prosperity, as institutions and governance structures greatly influence a nation's economic well-being~\cite{robinson2012nations}. Selecting the right government can foster enduring economic growth and development, impacting resource allocation, 
and citizen welfare~\cite{przeworski2000democracy}. 
On the other hand, political uncertainty stemming from unclear election outcomes can deter capital investment, thereby hindering economic development~\cite{political-economy-of-growth}. Thus, the stability and predictability of political environments, particularly during election cycles, are vital for attracting business investments and promoting economic progress~\cite{wlezien2002timeline}. 
Consequently, the need to forecast election results arises from the stakeholders' need to make well-informed choices. Businesses, investors, and politicians depend on election forecasts to predict policy shifts, evaluate market conditions, and manage risks \cite{goodell2020election}. Understanding potential election outcomes allows citizens to prepare for policy changes and potentially mobilize around issues they care about \cite{Bimber_2003}. Opinion polls, exit surveys and, more lately, social media studies try to fill this need by predicting election results. 

Social media sites like X (Twitter) allow people to voice their political opinions, and traditionally X's potential as a rich source of real-time, large-scale, and publicly available data for studying political behavior and sentiment has led to its growing use for election prediction. Prior works~\cite{gayo2013meta, Tumasjan_Sprenger_Sandner_Welpe_2010} have 
demonstrated the ability of social media to anticipate election outcomes accurately. 
Researchers have utilized 
tweet content, network architecture, and multi-modal signals to forecast election outcomes in several nations. Methods such as sentiment categorization, count-based techniques, and machine learning models such as VADER and BERT have played a crucial role in making these predictions~\cite{jungherr2016twitter, barbera2015tweeting}.

Although social media shows potential in forecasting election results, there are divergent opinions on its efficacy. Several studies indicate that social media might offer helpful information on public opinion and political trends \cite{conover2011political}. However, opposing views suggest that social media cannot accurately represent the complete range of voter behavior and can be impacted by biases and false information \cite{bond201261}, in addition to other challenges. To begin with, currently popular approaches often lack understanding of the general world. Additionally, the sentiment analysis methods frequently rely solely on keyword presence to identify the referenced political party. They assign a single sentiment score to an entire tweet. However, tweets often contain references to multiple parties within a single tweet. In cases where a tweet expresses positive sentiment towards one party and negative sentiment towards another, current methods fail to disambiguate the references to different parties within the text, 
due to the presence of mixed polarity within the sentences. This 
can lead the classifiers to generate a neutral score despite the presence of both positive and negative sentiment towards different political actors.

In recent times, Large language models (LLMs) have demonstrated an extraordinary ability to fathom enormous amount of 
world knowledge~\cite{
hendrycks2020measuring, zhong2023agieval}. The basic principle underlying the GPT models is to compress the world knowledge into the decoder-only Transformer model by language modeling, such that it can recover (or memorize) the semantics of world knowledge and serve as a general-purpose task solver \cite{zhao2023survey}. In this paper, we investigate whether LLMs can navigate the challenges with social media data, as mentioned earlier, and be used for predicting election outcomes without any fine-tuning in a zero-shot fashion. 

We specifically take the 2022 state assembly elections in India (Punjab and Uttar Pradesh) as the use case. We collected extensive longitudinal data from X alongside the opinion poll and exit poll results released by several psephology companies. Our experiments with two open source LLMs (Llama 2 and Zephyr) show that, compared to more traditional opinion polls and exit polls, LLM-based predictions can come much closer to the actual election outcomes. Overall, to our knowledge, ours is the first attempt to apply LLM for predicting election outcome and can offer a new perspective in this ever-expanding research domain.

\section{Literature Review}\label{section-Literature-Review}
One of the earliest studies in this domain was conducted by \cite{Tumasjan_Sprenger_Sandner_Welpe_2010}, who analyzed tweets related to the 2009 German federal election. They discovered that the volume of tweets mentioning a political party could serve as a plausible reflection of the vote share, with a predictive power comparable to traditional election polls. This approach was further refined by \cite{o2010tweets} through incorporating sentiment detection, which led to replicating consumer confidence and presidential job approval polls using data from X. Following these pioneering studies, researchers worldwide adopted the volume of tweets combined with automatic sentiment detection as the primary method for predicting election results. For instance, studies have been conducted in the Netherlands \cite{KLEINNIJENHUIS2013168}, France \cite{sokolova2018elections}, 
and the USA \cite{alashri2016analysis, gayo2011limits}. Additionally, novel approaches such as regression or time series methods \cite{o2010tweets, ceron2014every} and models using traditional polls for training or comparing results \cite{mancini2013new} were introduced.

In an attempt to enhance the predictive power of social media data, researchers have also explored the use of advanced machine learning algorithms. For instance, \cite{ElecBERT} developed the ElecBERT model, which leverages sentiment analysis of election-related tweets to predict electoral outcomes with high accuracy. Similarly, \cite{chauhan2021emergence} employed sentiment analysis methods to analyze X data during the 2019 Indian election, reportedly achieving predictions that closely mirrored the actual results.
Beyond sentiment analysis, multimodal approaches have been proposed to integrate various data types for election prediction. \cite{You_Luo_Jin_Yang_2015} presented a novel method that combined social multimedia data with a CNN-based visual sentiment analysis model to predict US election outcomes. \cite{huang2020attention} introduced an attention-based model that performs image-text sentiment analysis, further expanding the scope of data types used for election prediction. The role of geographic information in election prediction has also been investigated. \cite{9225511} proposed a location-based model integrating state-level sentiment analysis, offering a more granular approach to understanding regional political dynamics. Both \cite{9789178} and \cite{Rashed_Kutlu_Darwish_Elsayed_Bayrak_2021} assessed location-weighted counting methods in the context of the 2018 Turkish election, emphasizing the importance of spatial factors in electoral analysis. Additional techniques such as event analysis, social network analysis, and heterogeneous information diffusion have been employed to enhance the accuracy of election predictions \cite{xie2018big,li2019deep,Jiang_Ren_Ferrara_2023}. The computational complexity of these predictive models has been examined by \cite{baumeister2023complexity}, who investigated the trade-offs between model complexity and predictive power. Comprehensive surveys \cite{reveilhac2023influences, alvi2023frontiers} have provided valuable overviews of the field, identifying current trends and future research opportunities.
In conclusion, the use of social media data, particularly from X, has emerged as a powerful tool for predicting election results. Researchers have significantly enhanced the accuracy of election predictions by incorporating sentiment analysis, multimodal approaches, geographic information, and advanced machine learning algorithms. However, challenges related to data collection, model complexity, and computational resources remain, offering opportunities for further research in this rapidly evolving field.

\section{Background on the 2022 Punjab and UP Assembly Elections}
India, known for its vibrant democracy, conducts elections every 5 years at various levels of government, including the legislative assembly elections held in its states and union territories. The most distinguishing feature of Indian elections is the diversity of its electorate, which encompasses people from several cultural, linguistic, and socio-economic backgrounds. As per the statistical reports released by the Election Commission of India \cite{ElectionCommissionOfIndia_2022}, In 2022, Punjab, with a population of over 30 million, saw a very competitive election in which elections were contested in 117 constituencies. During the 2022 legislative assembly election, the state recorded 22.9 million eligible voters, comprising 11.6 million males and 11.3 million females. Notably, around 45\% of voters belonged to the 18-35 age group, highlighting the significant youth demographic in Punjab's electorate \cite{ElectionCommissionOfIndia_2022}. The election saw a diverse range of political parties participating, including the Indian National Congress (INC), Shiromani Akali Dal (SAD), and Aam Aadmi Party (AAP). The electoral contest in Punjab was marked by a robust candidate count, with hundreds of individuals vying for seats across the state. Uttar Pradesh (UP), India's most populous state with over 240 million people, featured a complex electoral landscape in the 2022 elections, with 403 constituencies. During the 2022 legislative assembly election, the state recorded 134.6 million eligible voters, comprising 70.4 million males and 64.2 million females \cite{ElectionCommissionOfIndia_2022}. UP's electorate displayed a broad demographic mix, reflecting the state's diverse social fabric. The electoral fray in UP was characterized by a multitude of political parties, including the Bharatiya Janata Party (BJP), Samajwadi Party (SP), Bahujan Samaj Party (BSP), and Congress. The state witnessed a significant number of candidates vying for seats, underlining the intense competition and strategic significance of Uttar Pradesh in the political landscape of India. The diverse demographics, constituency distribution, and political party participation in these states underscored the complexities of India's democratic process.

\begin{table*}
\centering
\caption{Search terms, hashtags, and user mentions used to extract data from X}
\label{tab:search-terms}
\begin{tabular}{lp{0.85\linewidth}}
\toprule
Party & Search queries \\
\midrule
BJP &  BJP, @BJP4India, @narendramodi, \#BJPwinningUP, @AmitShah, @myogiadityanath, yogi, @JPNadda, @BJP4TamilNadu, @BJP4UP, \#PunjabwelcomesModiji, \#Modigoback \\
\addlinespace
INC & @INCIndia, @RahulGandhi, Rahul Gandhi, @INCPunjab, channi, @priyankagandhi, \#111CongressDubara, @CHARANJITCHANNI, @INCUttarPradesh \\
\addlinespace
AAP & aam aadmi party, \#AAP, kejriwal, @AAPPunjab, @ArvindKejriwal, @AamAadmiParty, \#LokandaCM, \#ArvindKejriwal, \#KejriwalVsAll, @msisodia, \#AAPdePaap, @BhagwantMann \\
\addlinespace
SAD & @AkaliDal\_A, @Officeofbadal, @MSBADAL, @HarsimratBadal, @officialYAD, @bsmajithia, SGPC, Shiromani akali dal \\
\addlinespace
BSP & @Mayawati,@BSPIndia, bahujan samaj, mayawati, @AnandAkash\_BSP, @satishmisrabsp \\
\addlinespace
SP & @yadavakhilesh, samajwadi, @samajwadiparty, shivpal Yadav, \#MulayamSinghYadav \\
\addlinespace
General & \#PunjabElections2022, \#UPElections2022, \#UttarPradeshElections2022 \\
\bottomrule
\end{tabular}
\end{table*}

\section{Methodology}\label{section-methodology}
\subsection{Dataset collection}\label{ss-dataset-collection}
\subsubsection{Tweets}\label{sss-dataset-collection-tweets}
We collected tweets through the official X (formerly Twitter) API. To ensure the relevancy of the dataset and to capture the recent public sentiment, we collected tweets between February 1, 2022, and March 8, 2022, right before and during the elections, using search terms, hashtags, and user mentions. Table \ref{tab:search-terms} lists the search queries used for collecting data. Extracting data using these search terms, we collected 403,917 tweets,
\subsubsection{Opinion Polls and Exit Polls}\label{sss-dataset-collection-tweets}
Opinion polls and exit polls are conducted by various media houses prior to the elections to gauge public sentiment and try to predict the outcome of the elections. They publish these results on their websites and broadcast them during newsroom hours as debates. Both opinion polls and exit poll survey results for the election conducted by various reputed media houses as pollsters were curated and published by Oneindia on their website for Punjab \cite{Punjab-exit-opinion-poll} and Uttar Pradesh (UP) \cite{UP-exit-opinion-poll}, which we directly used in our analysis. 

\begin{figure*}[t]
    \centering
    \includegraphics[width=\textwidth]{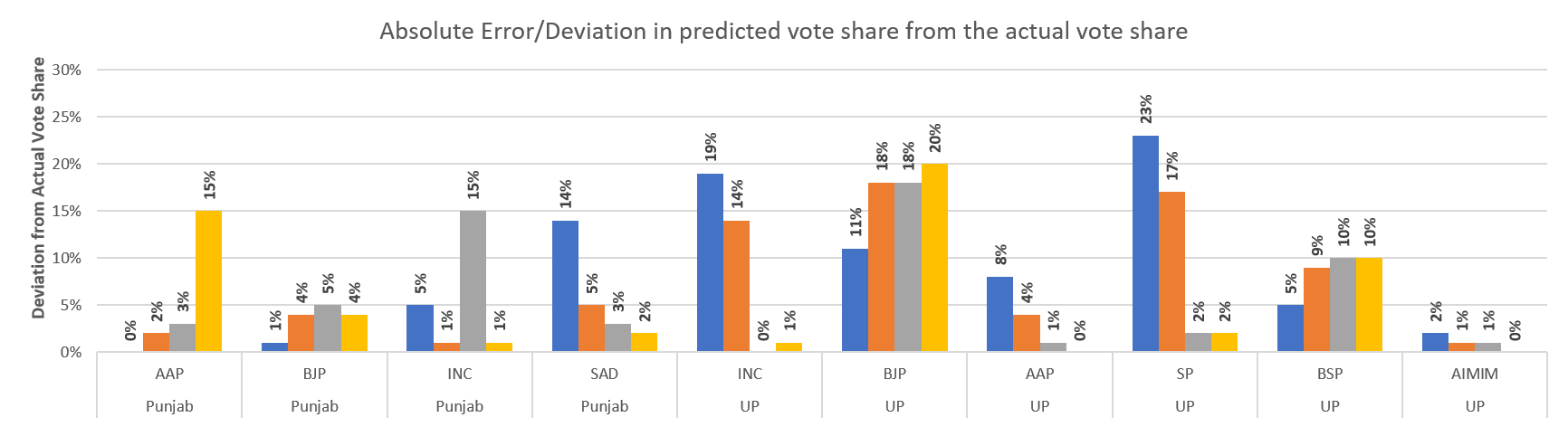}
    \caption{Absolute error/deviation of the predicted vote shares using LLMs and opinion polls and exit polls from the actual vote shares}
    \label{fig:deviation-llama-zephyr-opinion-exit-actual}
\end{figure*}

\textbf{Ethical Considerations} The data in this work has been collected only using Twitter API, respecting X’s terms of service. We exclusively collected the publicly available data, followed well-established ethical procedures for social data, and obtained approval from the ethics committee at the first author’s institution. 

\subsection{Data Preparation}\label{ss-data-prep}
\subsubsection{Tweets Dataset}\label{ss-data-prep-tweets}
In order to make full use of the language comprehension skills of the Large Language Models (LLMs), we deliberately refrained from carrying out any preprocessing on the tweets dataset, except for eliminating duplicate tweets. We dropped duplicate records on the combination of (``ID'', ``created\_at'', ``username'', and ``tweet'') columns. The final dataset, after the completion of cleanup operations, consisted of 241,301 tweets.

\subsubsection{Opinion and Exit Polls}\label{ss-data-prep-opinion-exit-survey}
As various pollsters released a predicted range of values for every party in their survey results, for every pollster, we computed the median value of the range predicted by them for a party and converted them to percentage share.

\subsection{Identify sentiment using LLM}\label{ss-LLM}
After data collection and cleanup, the next task was identifying which parties and states were referred to in a tweet and the associated sentiment score for each reference. We used two large language models- Zephyr-7B-$\beta$ \cite{tunstall2023zephyr} and Llama-2-13B \cite{touvron2023llama}. We chose these models because they are open, their inference can run on a low-cost consumer GPU, and at the time of experimentation, their output was known to be more accurate and coherent than other 7B and 13B parameter open foundation models.

\begin{figure*}[t]
    \centering
    \includegraphics[width=1.01\textwidth]{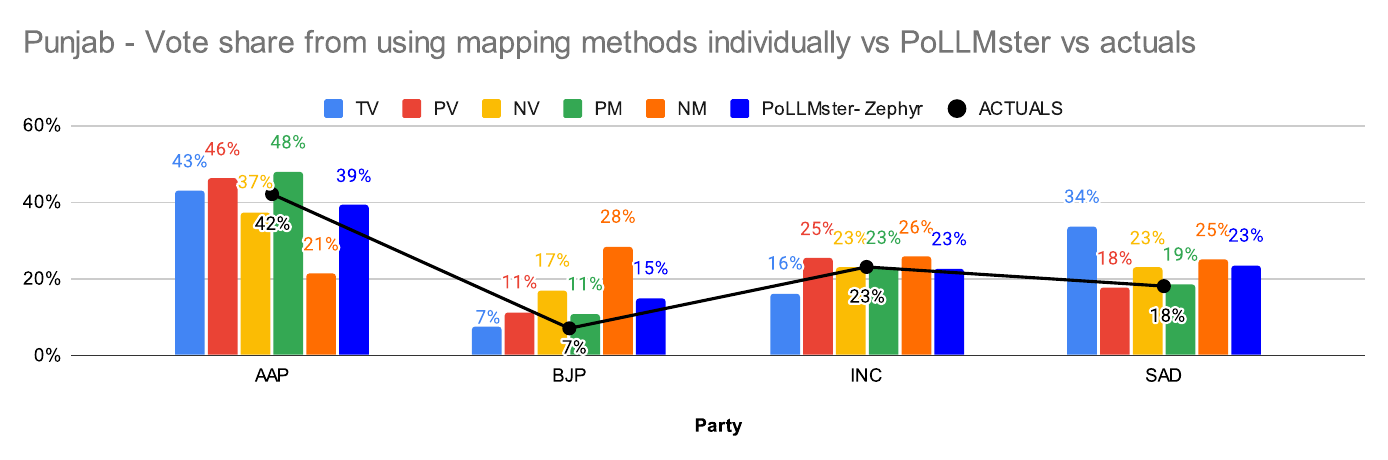}
    \caption{Predicted vote shares from using different mapping methods individually vs. PoLLMster (Zephyr7B-$\beta$) vs. the actual vote share for Punjab}
    \label{fig:zephyr-punjab-compare-methods}
\end{figure*}

The prompt given to Llama-2 was:\\
\textit{
\textless s\textgreater [INST]\textless\textless SYS\textgreater\textgreater\\
You are an expert political analyst specializing in political social media posts. Use all your knowledge from the Indian political landscape. Your task is to identify which Indian political party and state is being talked about in the following Tweet. Also, what is the sentiment score expressed in that reference of party and state? The hashtags, mentions, and names of people used in Tweet can help you identify the party and state. Your answer should be a JSON with three keys- party, state, and sentiment\_score. You have to choose only one among the following parties-- Bhartiya Janta Party (BJP), Congress (INC), Aam Aadmi Party (AAP), Shiromani Akali Dal (SAD), Samajwadi Party (SP), Bahujan Samaj Party (BSP). The state can be either Uttar Pradesh (UP) or Punjab. Sentiment scores can range between -1 (very negative) to +1 (very positive). Reply with only the JSON containing state, party, and sentiment\_score. Do not write anything other than that. Do not explain your answer.\\
\textless\textless /SYS\textgreater\textgreater \\ \#\#\# Instruction: Identify the party, state and sentiment score in the Tweet. Reply with only JSON. Here is the Tweet: \{tweet\} \\ \#\#\#Response:  [/INST]
}

The same tweets were processed through Zephyr as well, using the following prompt: \\
\textit{
\textless $\vert$system$\vert$\textgreater \\
You are an expert political analyst specializing in political social media posts. Use all your knowledge from the Indian political landscape. \textless/s\textgreater \\
\textless $\vert$user$\vert$\textgreater  \\
You are an expert Indian political analyst. You are given a tweet. The tweet may be in English, Hindi, Punjabi, or mixed code. It may directly or indirectly refer to one or more political parties. Your task is to identify which Indian political parties and states are being talked about in that Tweet, along with the sentiment score associated with that party. The hashtags, mentions, and names of people used in Tweet, along with all the knowledge you have about Indian elections, will provide you enough context to identify the party and the state. A Tweet may refer to more than one party. \\
Your answer should be a JSON array, each object containing three keys- party name, state, and sentiment score. If a tweet talks about more than one party, then your response should contain the party name, sentiment score, and state for each of the identified parties. 
You need to adhere to the following constraints-- For the party, only the following values are allowed-``Bhartiya Janta Party (BJP)'', ``Congress (INC)'', ``Aam Aadmi Party (AAP)'', ``Shiromani Akali Dal (SAD)'', ``Samajwadi Party (SP)'', ``Bahujan Samaj Party (BSP)''. Allowed values for the State are- ``Uttar Pradesh (UP)'' and ``Punjab''. Sentiment score can range between -1 (very negative) to 1 (very positive). \\ 
Reply with only the valid JSON containing state, party, and sentiment score. Do not write anything other than JSON. Do not explain your answer. Your output should be in the format: 
[\{``party'': ``\textless party name\textgreater'', ``state'':``\textless state name\textgreater'', ``sentiment\_score'':\textless sentiment score\textgreater\}]. 
\\
Here is the Tweet:  ``\{tweet\}'' \\ \textless /s\textgreater 
\textless$\vert$assistant$\vert$\textgreater
}

\subsection{Post-processing} \label{ss-postprocessing}

While we had used an LLM output format enforcer to compel the output as JSON, at the time of experimentation, the format enforcers were not 100\% effective, as in many cases, LLM has output additional text such as a context in the beginning or an explanation of its answer towards the end of its response. Hence, cleaning up the LLM outputs to remove any redundant data it had produced was imperative. We implemented a three-step process for extracting the relevant information.
\begin{enumerate}
    \item If the output of LLM was valid JSON, we parsed the state, party, and sentiment scores. 
    \item If the raw output of LLM was not a valid JSON, we used regular expressions to extract only the JSON component from the LLM output. We tried parsing it to extract the state, party, and the associated sentiment score. 
    \item If the regular expression patterns used in the previous step produced no matches, we attempted to extract the tokens that immediately followed ``state'', ``party'', and ``sentiment\_score'' in the LLM output to extract the state, party, and sentiment score, respectively. 
\end{enumerate}
We observed that the outputs obtained from the LLM had multiple aliases for the same state and party. For example, for different tweets mapped to the state of Uttar Pradesh, the LLM output the state as ``Uttar Pradesh", ``UP," and ``Uttar Pradesh (UP)" and at times also returned the names of cities in the state instead of the name of the state itself, such as Jaunpur, Amethi, and Varanasi. To fix this problem, we identified the aliases representing 99.9\% of the records through cumulative distribution tables and mapped them to a single name. Similarly, we repeated the same process to map various aliases of the same party to a distinct party name.

\begin{table*}[t]
\centering
\caption{Vote shares — Predicted vs. Actual.}
\label{tab:voteshare-prec}
\begin{tabular}{@{}ccccccc@{}}
\toprule
\multirow{2}{*}{\textbf{State}} &
  \multirow{2}{*}{\textbf{Party}} &
  \multicolumn{2}{c}{\textbf{LLM Predicted Vote Share}} &
  \multicolumn{2}{c}{\textbf{Predicted through surveys}} &
  \multirow{2}{*}{\textbf{Actual Vote Share}} \\ \cmidrule(lr){3-6}
 &
   &
  \textbf{Llama-2} &
  \textbf{Zephyr} &
  \textbf{Opinion Poll} &
  \textbf{Exit Poll} &
   \\ \midrule
Punjab & AAP   & 42\% & 44\% & 45\% & 57\% & 42\% \\
Punjab & BJP   & 8\%  & 11\% & 2\%  & 3\%  & 7\%  \\
Punjab & INC   & 18\% & 22\% & 38\% & 24\% & 23\% \\
Punjab & SAD   & 32\% & 23\% & 15\% & 16\% & 18\% \\
UP     & INC   & 21\% & 16\% & 2\%  & 1\%  & 2\%  \\
UP     & BJP   & 52\% & 59\% & 59\% & 61\% & 41\% \\
UP     & AAP   & 8\%  & 4\%  & 1\%  & 0\%  & 0\%  \\
UP     & SP    & 9\%  & 15\% & 34\% & 34\% & 32\% \\
UP     & BSP   & 8\%  & 4\%  & 3\%  & 3\%  & 13\% \\
UP     & AIMIM & 2\%  & 1\%  & 1\%  & 0\%  & 0\%  \\ \bottomrule
\end{tabular}
\end{table*}

\begin{figure*}[t]
    \centering
    \includegraphics[width=1.01\textwidth]{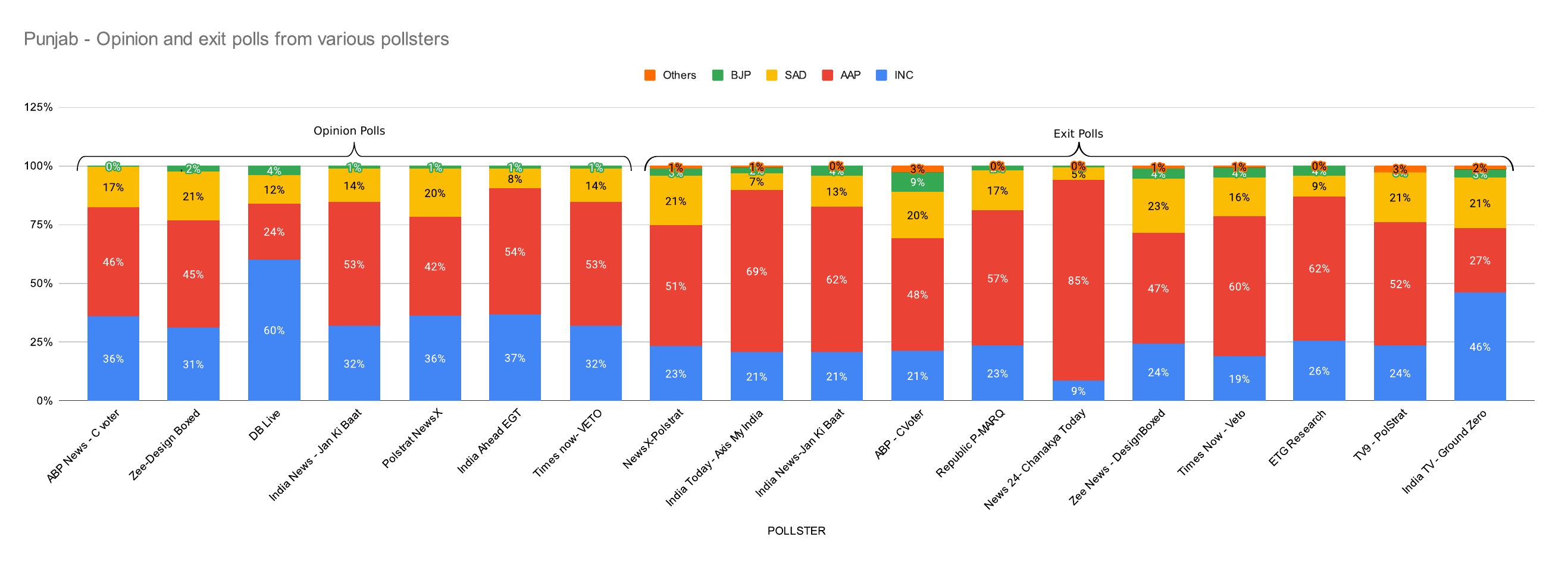}
    \caption{Punjab - Exit and opinion polls from various pollsters}
    \label{fig:punjab-opinion-exit-pollsters}
\end{figure*}

\begin{figure*}[ht]
    \centering
    \vspace*{-10mm}
    \includegraphics[width=1.01\textwidth]{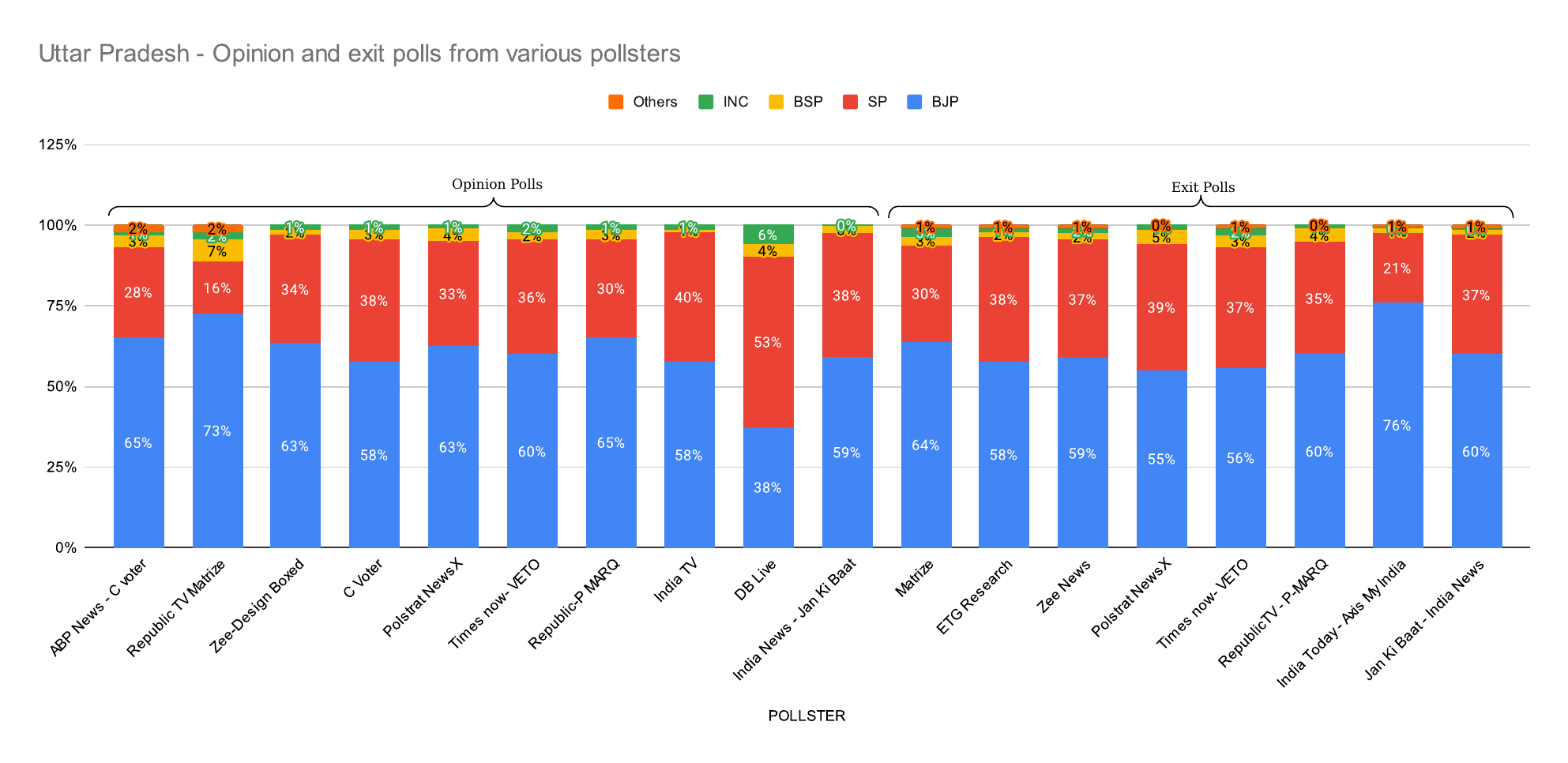}
    \caption{Uttar Pradesh - Exit and opinion polls from various pollsters}
    \label{fig:up-opinion-exit-pollsters}
\end{figure*}

\subsection{Mapping sentiment scores to vote shares}\label{ss-mapping}
After computing the sentiment score for each instance of party and state detected in a Tweet, the resulting data were consolidated to determine the final vote share for the competing parties in the two states. Numerous previous studies, including \cite{Tumasjan_Sprenger_Sandner_Welpe_2010}, used the volume of tweets made for a party as a proxy for their vote share. However, sentimentality in the tweets should also be taken into account. While certain tweets express favorable sentiments towards a particular political party, others are critical in nature and are composed of supporters of that party critiquing an opposing party. The sentiment expressed in tweets can reflect the prevailing mood of the electorate, potentially influencing the degree of incumbency or anti-incumbency towards a political party; that is, in addition to the sentiment expressed in tweets favoring a party, the sentiment score of tweets criticizing the competitor or opposing parties should also have a direct bearing on the vote share of a given political party. Consequently, we used five different functions in Eqn. \ref{eqn:vote-share-mapping-methods} to compute vote share from the sentiment scores for all parties in the two states of UP and Punjab using the two models- Llama-2-13B and Zephyr-7B-$\beta$. 

\begin{equation} \label{eqn:vote-share-mapping-methods}
    \begin{aligned}
    &VS_{p}^{PM} = \frac{PM_{p}}{ \sum_{i} PM_{i} } & 
    &VS_{p}^{NM} = \frac{\sum_{i' \neq i} NM_{i'}}{ \sum_{i} NM_{i} } \\
    &VS_{p}^{PV} = \frac{PV_{p}}{ \sum_{i} PV_{i} } &
    &VS_{p}^{NV} = \frac{\sum_{i' \neq i} NV_{i'}}{ \sum_{i} NV_{i} } \\
    &VS_{p}^{TV} = \frac{TV_{p}}{ \sum_{i} TV_{i} } & 
    \end{aligned}
\end{equation}
where, \\
$VS_p$: Vote share for party $p$  \\
$PM$: Sum of the magnitude of positive sentiment scores \\
$NM$: Sum of the magnitude of negative sentiment scores of other parties \\
$PV$: Volume of tweet references with a positive sentiment  \\
$NV$: Volume of tweet references with a negative sentiment for other parties \\
$TV$: Total references of a party in tweets by volume
\\
The vote shares obtained using these mapping methods are listed in Table \ref{tab:vote-share-methods-models}. We then computed \textbf{PoLLMster vote share}-- the final predicted vote share, as the linear average of the vote shares obtained by the five mapping methods from Eqn. \ref{eqn:vote-share-mapping-methods}, as shown in Eqn. \ref{eqn:vote-share-combined-equation}.\\
$\textbf{PoLLMster vote share VS(p):}$
\begin{equation} \label{eqn:vote-share-combined-equation}
    \begin{aligned}
    VS (p) = & \frac{1}{5}\sum_{i=1}^5{VS_p ^i}  \\ 
    \end{aligned}
\end{equation}

\section{Experimental Evaluation}\label{section-results} 
We tested our methodology to predict the vote share of all the parties contesting in Punjab and Uttar Pradesh (UP). The final vote share computed using our methodology is available in table \ref{tab:voteshare-prec}. As our Tweets dataset was limited to a few days just before the poll, it is more logical to compare the LLM-predicted results with the opinion polls results than comparing it with the exit poll results as the latter is conducted after the polls, and so is expected to be more accurate. However, our method of using LLMs to forecast the election results outperforms both- opinion and exit poll results for most states and parties, closely matching the actuals.

\subsection{Vote share mapping methods}
An analysis revealed that the sentiment-to-vote-share mapping methodologies produced inconsistent outcomes when applied in isolation to predict the vote share. Fig. \ref{fig:zephyr-punjab-compare-methods} compares the predicted vote shares obtained for Punjab using different mapping methods in isolation vs. PoLLMster. We obtained similar results using Llama in Punjab and UP, which are not included here for brevity. In general, it was found that mapping methods based on the positive sentiment tweets- $VS_p^{PV}$ and $VS_p^{PM}$ performed better than those based on the negative sentiment tweets- $VS_p^{NV}$ and $VS_p^{NM}$. However, this was not always true. For instance, in Fig. \ref{fig:zephyr-punjab-compare-methods}, for the AAP and the BJP, predictions from $VS_p^{PV}$ and $VS_p^{PM}$ were closer to the actual as compared to the ones from $VS_p^{NV}$ and $VS_p{NM}$. On the contrary, the INC's vote share predicted from $VS_p^{NV}$ (23\%) was better than the vote share predicted from $VS_p^{PV}$ (25\%) as the actual vote share was 23\%. However, combining the vote share mapping methods by \ref{eqn:vote-share-combined-equation} evens the differences and results in the final predicted vote share close to the actual. By taking an average of the vote shares predicted from the five mapping methods, it can be observed that the PoLLMster vote share is consistently close to the actual vote share across parties in both states compared to using each method individually, irrespective of the LLM used. The final predicted vote share results using Llama-2 and Zephyr are available in Table \ref{tab:voteshare-prec}. As observed in Fig. \ref{fig:deviation-llama-zephyr-opinion-exit-actual}, the error in predicting the vote share using the two LLMs is similar to that forecasted by pollsters in their opinion and exit polls survey reports.

\subsection{Stability of the results}
While Llama-2 and Zephyr predicted the winning parties in both states accurately, it can be argued that the vote shares obtained from Llama-2 and Zephyr vary slightly, as seen in table \ref{tab:voteshare-prec}. The difference among the vote shares predicted by the two models was highest for SAD in Punjab at 9\%, with the average deviation being 5\%. While there is a difference in the vote shares predicted by the two models, the deviation is much lower than that of the opinion and exit poll-predicted results. The percentage share of each party from the opinion and exit polls, calculated in section \ref{ss-data-prep-opinion-exit-survey}, is shown for Punjab and Uttar Pradesh in Fig. \ref{fig:punjab-opinion-exit-pollsters} and Fig. \ref{fig:up-opinion-exit-pollsters} respectively. It can be observed that the survey results can vary significantly among different pollsters. For instance, in Fig. \ref{fig:up-opinion-exit-pollsters}, it can be seen that while DB Live forecasted that the Samajwadi Party (SP) would secure 53\% share, Republic TV's forecast for the party stood at a paltry 16\%. similarly, in Punjab, while News24 and India Today forecasted a whooping 85\% and 69\% share for the Aam Aadmi Party (AAP), India TV and DB Live predicted the party would secure a mere 27\% and 24\%, respectively. To conclude, although the two models forecast different vote shares, the difference is far smaller than the variation observed in the opinion and exit poll predictions.

\section{Conclusion and Discussion}\label{section-results}
The objective of this study was to forecast the results of the 2022 assembly elections in Punjab and Uttar Pradesh by employing computational social science techniques. Our approach used sophisticated natural language processing techniques with extensive linguistic models (LLMs) to examine the sentiment associated with political parties and candidates in X tweets. The results of our study show a strong link between the emotion ratings obtained from LLM outputs and the actual election outcomes. The sentiment research in Punjab successfully forecasted the triumph of the Aam Aadmi Party (AAP), as evidenced by their overwhelming majority. Similarly, in Uttar Pradesh, the sentiment ratings aligned with the Bharatiya Janata Party's (BJP) successful re-election, demonstrating the effectiveness of our prediction methodology. The analysis of these findings emphasizes the capacity of LLMs to comprehend and predict voter conduct. The precision of our forecasts highlights the significance of computational techniques in political science, especially in examining extensive social media data to assess popular sentiment. 
\subsection{Current limitations, biases and ethical concerns}
Several inherent challenges impact the precision of the predictions presented earlier. The low internet penetration rate in India, coupled with the diverse demographic and linguistic landscape, especially in Uttar Pradesh, limits the representativeness of X data. The demographic and linguistic diversity of the population further complicates the data collection process, as the sentiments expressed on X may not accurately mirror the electorate’s views. The incumbency advantage often results in a skewed perception of performance and increased voter support, potentially influencing the election outcome. Additionally, the propensity of X users to express political opinions does not necessarily reflect the broader voting population. This bias in data can lead to skewed results, as LLMs are known to learn, perpetuate, and amplify harmful social biases if not properly mitigated \cite{gallegos2023bias}. When applied to election prediction, large language models (LLMs) raise ethical concerns due to the possibility of manipulation and bias. These biases may originate from the data utilized in the training of LLMs, which could be influenced by societal prejudices or distorted by particular online communities.  Consequently, LLM forecasts may exhibit a bias towards particular candidates or deter engagement from specific demographic groups, thereby subverting the integrity of the democratic process. 
\subsection{Future work}
Despite already performing at par or better than the opinion and exit poll results, our results can be further enhanced in several ways, including using a more extensive dataset, better models, and prompt engineering. Increasing the volume of tweets collected could provide a more comprehensive dataset, reflecting a wider range of political sentiments. Employing more advanced proprietary models, such as GPT-4, Claude-3 and Gemini-Pro, may yield superior results due to their enhanced processing capabilities and understanding of nuanced language, as they currently top several LLM Leaderboards such as \cite{open-llm-leaderboard}. Furthermore, leveraging prompt engineering techniques and optimizing prompts could lead to more precise sentiment extraction and party identification. Being an active area of research, several works present varying techniques, including but not limited to exploiting patterns in prompts, automated prompt generation, chain-of-thought reasoning, and zero-shot reasoning, to make the model responses more accurate, free from any biases, and harmless \cite{brown2020language, radford2019language}. It is crucial to acknowledge that while LLMs can exhibit cognitive biases, there is cautious optimism about their performance, provided genuine biases are identified and efforts are made to reduce them  \cite{thorstad2023cognitive}.


In conclusion, while the current methodology may have limitations, it establishes a solid foundation for future exploration in the field. The application of LLMs in election forecasting is an emerging area of research that holds considerable promise. With ongoing advancements in computational models, data collection methods, and analytical techniques, there is potential for significant improvements in the accuracy of election predictions. This study serves as a catalyst for further investigation, highlighting the need for continuous innovation to address the challenges identified and to harness the full potential of LLMs in political analysis.

\bibliographystyle{IEEEtran}
\bibliography{main}
\end{document}